\documentclass{SCGE}

\begin{document}

\begin{picture}(0,0){\rm
\put(0,-39){\makebox[160truemm][l]{\bf {\sanhao\raisebox{2pt}{.}}
Article {\sanhao\raisebox{1.5pt}{.}}}}}
\put(0,-52){\jiuwuhao {\textcolor[rgb]{0.5,0.5,0.5}{\sf 
}}}
\end{picture}

\def\bm{\boldsymbol}

\def\dl{\displaystyle}
\def\du{\end{document}}
\def\pi{{\uppi}}

\Year{2013} %
\Month{??} %
\Vol{??} %
\No{??} %
\BeginPage{1} %
\EndPage{??} %
\AuthorMark{{\rm A.Y. Yang}, et al.}
\DOI{??} 

\title{A new method to analyse pulsar nulling phenomenon}

\author[1,2]{A.~Y. Yang}{}%
\author[3*]{J.~L. Han}{}
\author[1]{Na Wang}{}

\address[{\rm1}]{Xinjiang Astronomical Observatory, Chinese Academy of
  Sciences, 150 Science-1 street, Urumqi, Xinjiang, 830011, China}
\address[{\rm2}]{The University of Chinese Academy of Sciences,
  Beijing 100049, China} 
\address[{\rm3}]{National Astronomical Observatories, Chinese Academy
  of Sciences, 20A Datun Road, Chaoyang District, Beijing 100012,
  China}

\maketitle \vspace{-3.5mm}{\footnotesize\begin{center} 
Received August 29th, 2013; accepted September xx, 2013
\end{center}}\vspace*{-5mm}

\begin{center}
\rule{16.5cm}{0.4pt}
\parbox{16.5cm}
{\begin{abstract} 
Pulsar nulling is a phenomenon of sudden cessation of pulse emission
for a number of periods. The nulling fraction was often used to
characterize the phenomenon. We propose a new method to analyse pulsar
nulling phenomenon, by involving two key parameters, the nulling
degree, $\chi$, which is defined as the angle in a rectangular
coordinates for the numbers of nulling periods and bursting periods,
and the nulling scale, $ N $, which is defined as the effective length
of the consecutive nulling periods and bursting periods. The nulling
degree $\chi$ can be calculated by $\tan \chi = N_{\rm nulling} /
N_{\rm bursting} $ and the mean is related to the nulling fraction,
while the nulling scale, $ N $, is also a newly defined fundamental
parameter which indicates how often the nulling occurs. We determined
the distributions of $\chi$ and $ N $ for 10 pulsars by using the data
in literature. We found that the nulling degree $\chi$ indicates the
relative length of nulling to that of bursting, and the nulling scale
$ N $ is found to be related to the derivative of rotation frequency
and hence the loss rate of rotational energy of pulsars. Their
deviations reflect the randomness of the nulling process.
\end{abstract}}
\end{center}\vspace*{-0.6cm}

\begin{center}
\parbox{16.5cm}
{\bf\jiuhao Key words: Pulsars; Radio sources; Radiation mechanisms}
\end{center}

\begin{center}
{}
\vspace*{-1mm}
\Cit{A.Y. Yang et al., A new method to analyse pulsar nulling phenomenon.
 Sci China-Phys Mech Astron, 2013, 55: 1--??, doi: ?? } 
\end{center}

\wuhao\vspace*{1.5mm}

\begin{multicols}{2}

\renewcommand{\baselinestretch}{1.08} \baselineskip 12.2pt\parindent=10.8pt

\renewcommand{\thefootnote}

\sec{1\quad Introduction}

\no Pulsar nulling is abrupt cessation in pulsar emission, as first
detected in [1], which is still not well understood. It often appears
as a broad band phenomenon of pulsar radio emission. Currently, among
2267 known pulsars, nulling phenomenon has been detected for 185
strong pulsars [1-22].

There have been many observations and analyses for the nulling
phenomenon for a number of strong pulsars [1-22].  The nulling
fraction ($NF$), which is the time fraction for the null emission
state, is often used to characterize the nulling properties of
pulsars. It is in general less than $10\%$ for most pulsars, but
rarely as high as more than $90\%$ for a few pulsars [5,12]. The null
duration is a range of one period to hundreds of periods for various
pulsars. The Rotating Radio Transient (RRAT) may be the nulling with
an extremely high nulling fraction, which emits one pulse in many
periods [3].
\vspace*{1mm}
\noindent\rule{2.5cm}{0.4pt} \\[0.1mm]{\qihao *Corresponding author (email:
hjl@nao.cas.cn)
}
%

Observations of individual pulses provide valuable hints for pulsar
emission mechanism and propagation process. The switches of pulsar
emission modes or magnetosphere states have been observed in many
pulsars [4-5], and nulling may regard as being an extreme form of such
switches. From a few pulsars, subpulse drifting rate [6-7] and
interpulse appearance [8] have been shown to be related to
nulling. The subbeam carousel model can be used to explain the
quasi-periodicity of nullings detected from some pulsars [9-11]. PSR
J0941$-$39 [3] switches not only between two states for nulling and
bursting, but also emits rare pulses during the nulling state as if it
is a RRAT.

\begin{table*} 
\centering
\caption{\normalsize Parameters and observational details for 10
  nulling pulsars in literature} 
{\small
  \begin{tabular}{lccrllll}
  \toprule[0.65pt]
   PSR Jname & PSR Bname & P    &  DM     & Obs.~Freq. & BW & Telescope    & Ref. \\
      &      &(s)     &(${\rm cm}^{-3}{\rm pc}$)&  (MHz)                   &  (MHz)      &             &\\
   \hline
 J0820$-$4114&B0818$-$41&0.545 & 113.4  & 325/610    &  16    &GMRT      & [19]     \\ 
 J1049$-$5833&-        &2.202 & 446.8   & 1518       &  576   &Parkes    & [5]    \\ 
 J1502$-$5653& -        &0.536 & 194.0  & 1374/1518  &288/576 &Parkes    & [2,5] \\ 
 J1701$-$3726&B1658$-$37&2.455 & 303.4  & 1518       &  576   &Parkes    & [5]    \\ 
 J1703$-$4851& -        &1.396 & 150.3  & 1518       &  576   &Parkes    & [5]    \\ 
 J1727$-$2739& -        &1.293 & 147.0  & 1518       &  576   &Parkes    & [5]    \\ 
 J1752$+$2359& -        &0.409 & 36.0   & 430        &  8     &Arecibo   & [20]    \\ 
 J1819$+$1305&-         &1.060 & 64.9   & 327        &  25    &Arecibo   & [11]    \\ 
 J1820$-$0509& -        &0.337 & 104.0  & 1518       &  576   &Parkes    & [5]    \\ 
 J1946$+$1805&B1944$+$17&0.441 & 16.2   & 430        &  8     &Arecibo   & [21]     \\ 
 \bottomrule[0.65pt]
\end{tabular}
}
\end{table*}

Previous studies of nulling pulsars have been focused on the
correlations between pulsar parameters and the nulling fraction
[12,13].  The most often discussed but without consensus is the
possible relation between pulsar characteristic age and nulling
fraction [22,5,12,13]. However, it is not clear if the nullings occur
randomly or periodically [11,14], depending on pulsars. The {\it
  Wald-Wolfowitz runs test} was used to analyze the randomness of
pulsar nulling [14,18]. The Markov process was recently used by Cordes
[17] to understand the transitions between nulling and bursting
states.

In this paper, we propose a new method to analyse pulsar nulling
phenomenon, by involving two key parameters, the nulling degree,
$\chi$, which is defined as the average angle in a rectangular
coordinates for the numbers of nulling periods and bursting periods,
and the nulling scale, $N$, which is defined as the effective length
of the consecutive nulling periods and bursting periods.  We believe
that the distributions of these two parameters can characterize the
nulling phenomenon with a deeper insight in emission processes than
the often-used nulling fraction. We obtain data from literature for 10
pulsars and analyse for these two parameters, and discuss the possible
relationship between the distribution of the two parameters and other
pulsar observational properties.

\sec{2\quad Nulling degree and scale for 10 pulsars}

Observations for pulsar nulling in general were made for significant
detections of single pulses of strong pulsars by using a sensitive
radio telescope. In order to get sufficient signal to noise ratio,
single-pulse sequences sometimes are integrated to short
subintegrations [5,18]. The nulling state is determined by the
threshold of three or five times of the rms in the on-pulse window for
detection of pulse emission. The nulling length ($N_{\rm nulling}$) is
counted by continuous pulsar periods when the emitted pulse energy is
below the threshold, and the bursting length ($N_{\rm bursting}$) is
the number of periods for continuously detected pulse emission above
the threshold.

In order to characterize the nulling properties properly, we count 
the $N_{\rm nulling}$ and $N_{\rm bursting}$ from the observed single-pulse
sequences, and plot in Fig.1 the numbers for any adjacent states.
We then define two parameters $N$ and $\chi$ as
\begin{figure}[H]
\centering
\includegraphics[width=0.35\textwidth,angle=-90]{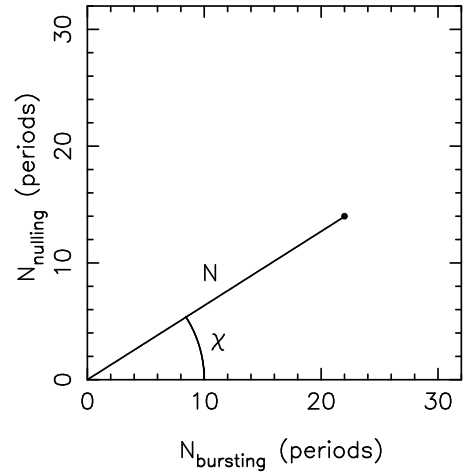}\vspace{-2mm}
\caption{Schematic view for the nulling scale $N$ and nulling degree
  $\chi$, derived from only one set of $N_{\rm bursting}$ and $N_{\rm
    nulling}$ for adjacent bursting and nulling. The distribution of
  $\chi$ indicates the degree of nulling, $ \langle \chi \rangle =
  0^o$ for no nulling, and $ \langle \chi \rangle = 90^o$ means very
  nulling such as a RRAT. Statistics of $ N $ indicates how often the
  nulling and bursting states switch.  }
\label{fig1}
\end{figure}
\begin{equation}
N =\sqrt{{N_{\rm nulling}}^{2}+{N_{\rm bursting}}^{2}}
\end{equation}
\begin{equation}
\chi =\arctan (\frac{{N}_{\rm nulling}}{{N}_{\rm bursting}})
\end{equation}
\no For a sequence of pulses, we obtain a distribution of data points
from a set of $N_{\rm nulling}$ and $N_{\rm bursting}$. The statistics
of these data distribution presumably describe the nulling properties
not only for the often-used nulling fraction via $\langle \chi
\rangle$, but also the nulling scale via $\langle N \rangle$.

\begin{figure*}
  \begin{tabular}{ccc}
    \includegraphics[width = 0.30\textwidth, height=0.30\textwidth, angle = -90]{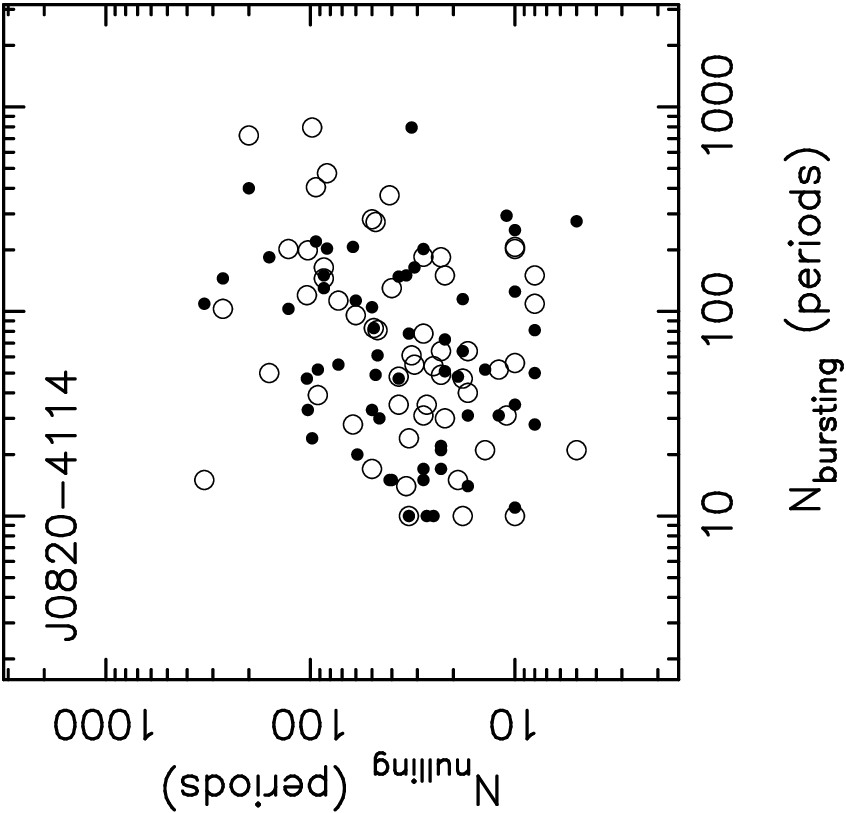}&
    \includegraphics[width = 0.30\textwidth, height=0.30\textwidth, angle = -90]{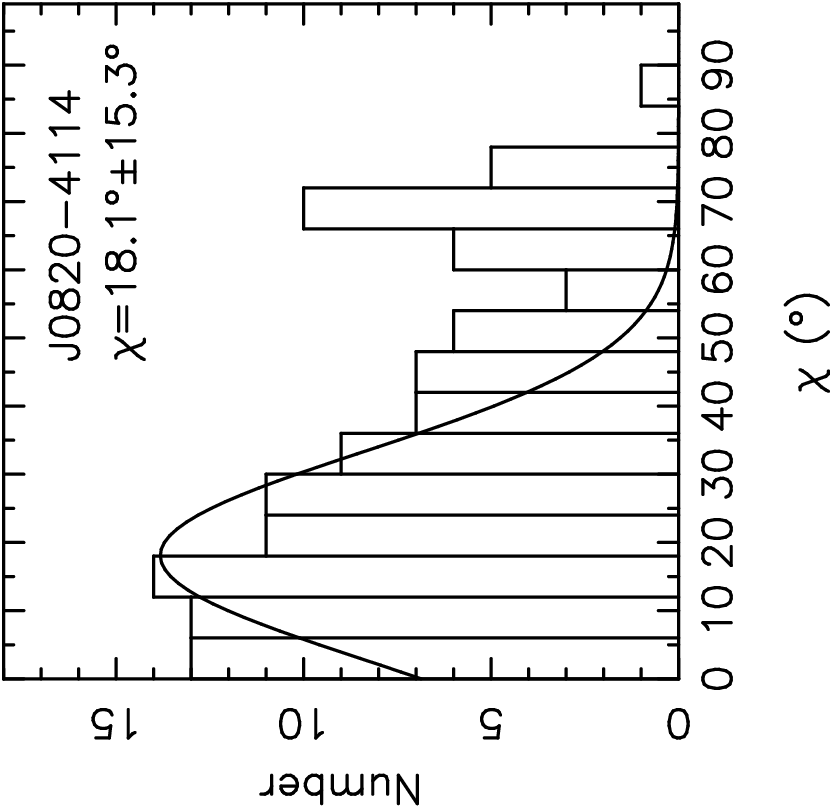}&
    \includegraphics[width = 0.30\textwidth, height=0.30\textwidth, angle = -90]{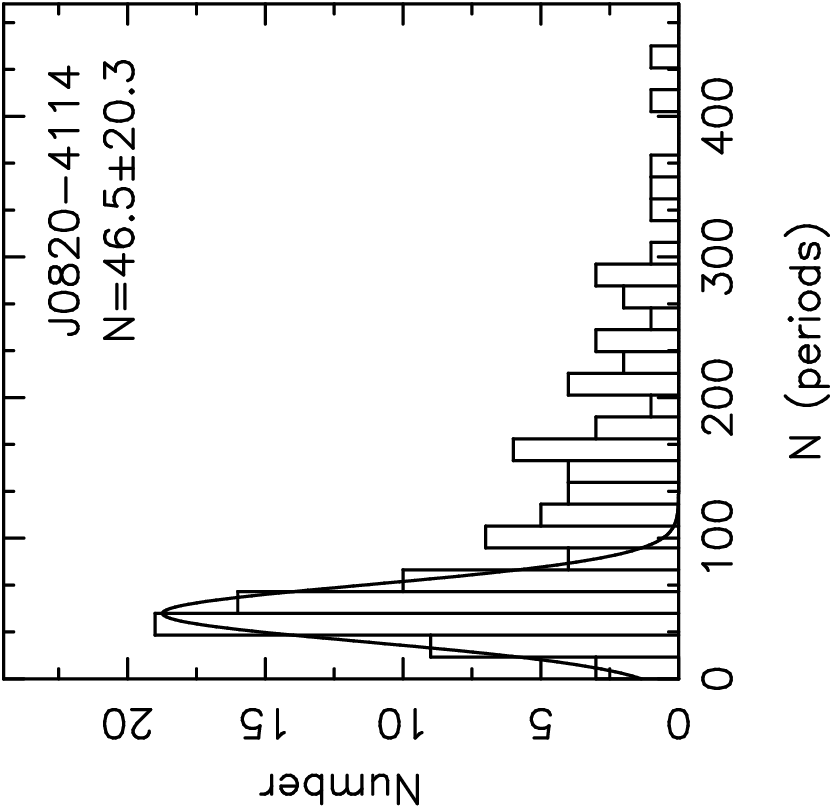}\\[2mm]

    \includegraphics[width = 0.30\textwidth, height=0.30\textwidth, angle = -90]{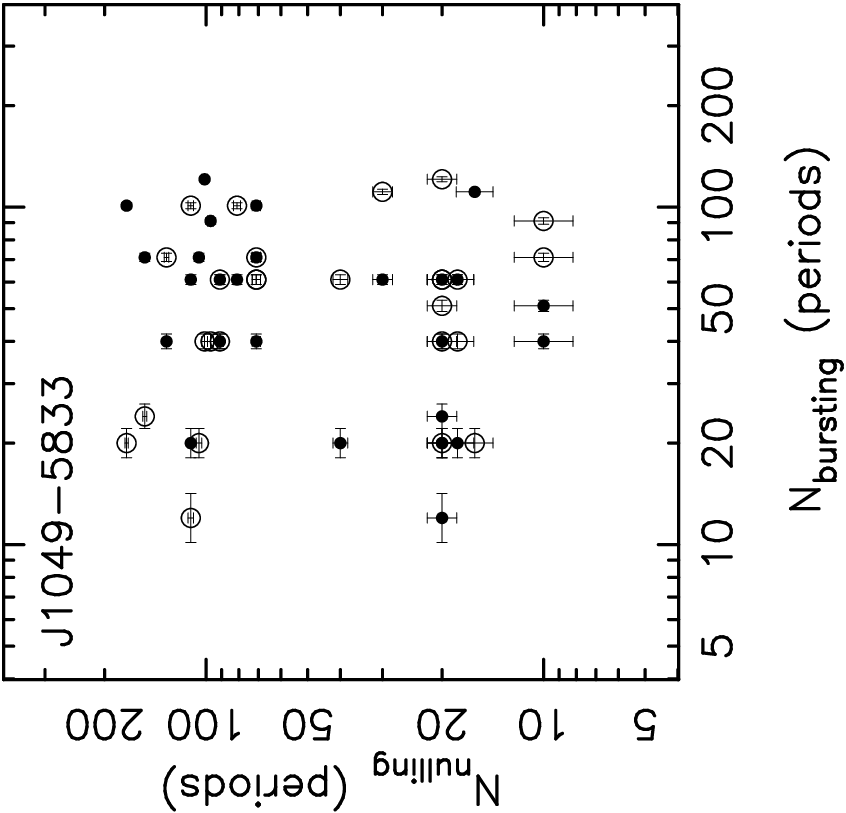}&
    \includegraphics[width = 0.30\textwidth, height=0.30\textwidth, angle = -90]{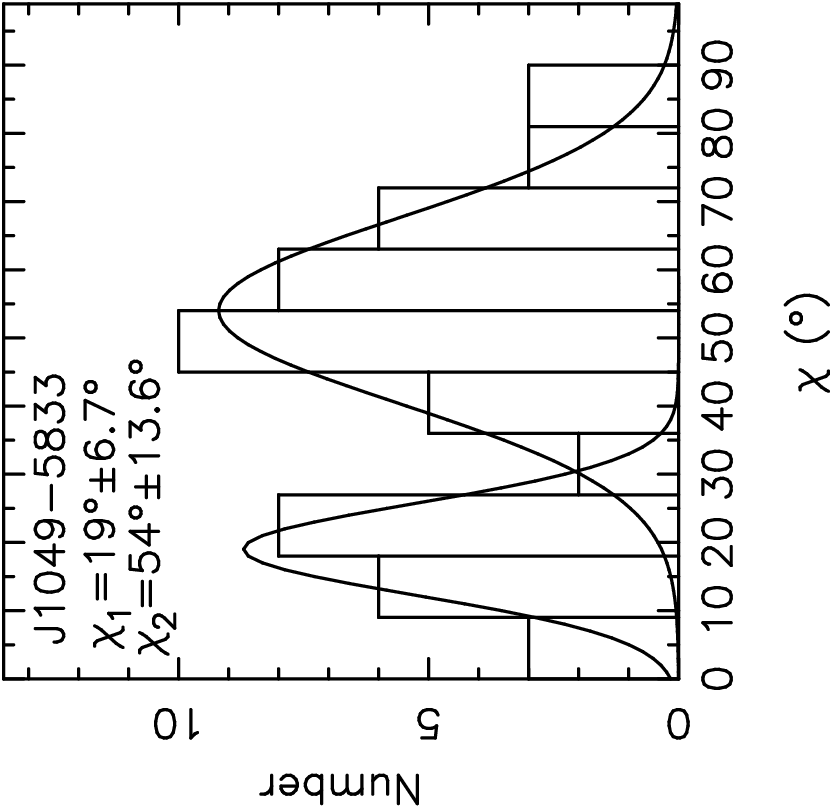}&
    \includegraphics[width = 0.30\textwidth, height=0.30\textwidth, angle = -90]{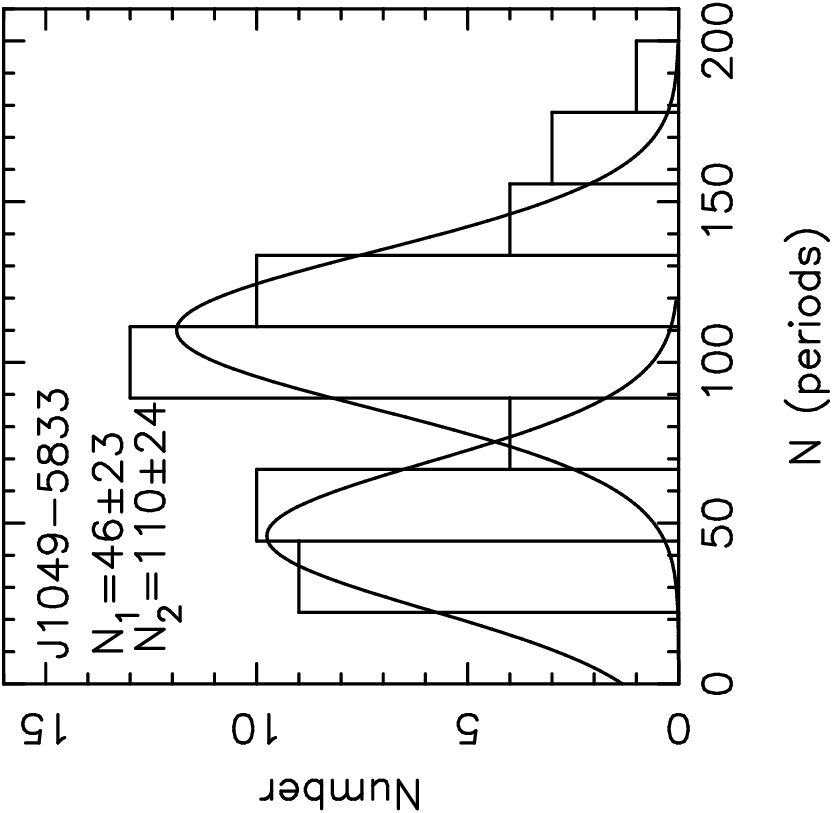}\\
    \includegraphics[width = 0.30\textwidth, height=0.30\textwidth, angle = -90]{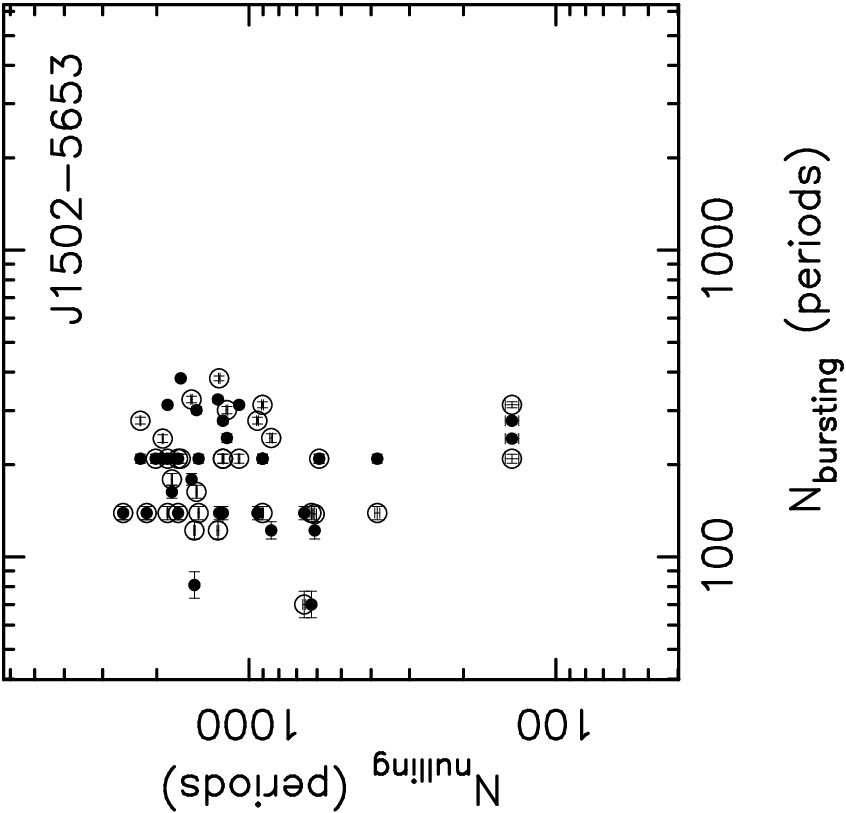}&
    \includegraphics[width = 0.30\textwidth, height=0.30\textwidth, angle = -90]{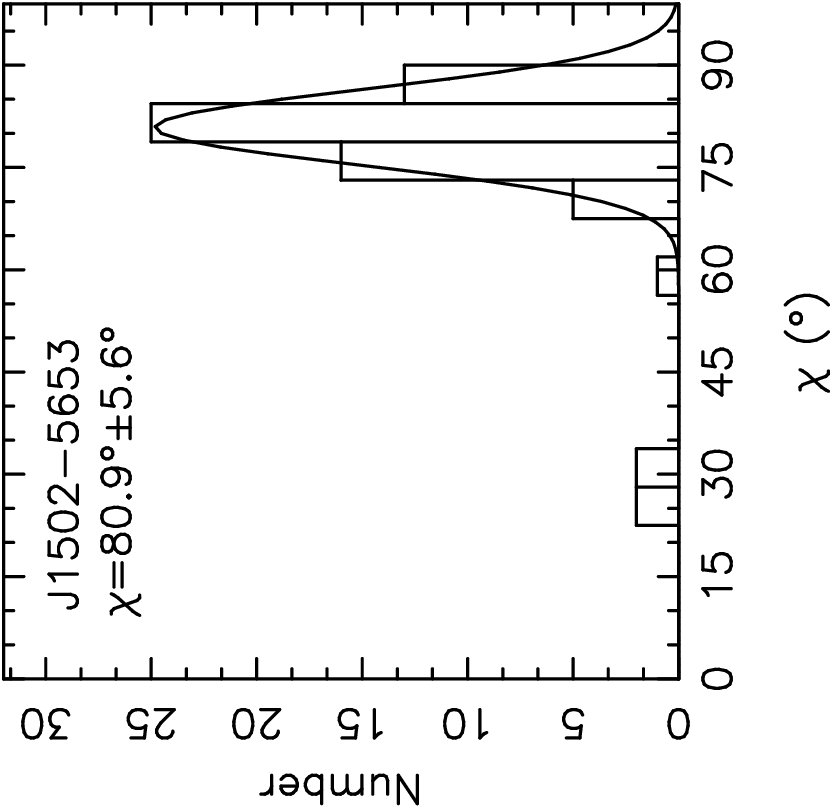}&
    \includegraphics[width = 0.30\textwidth, height=0.30\textwidth, angle = -90]{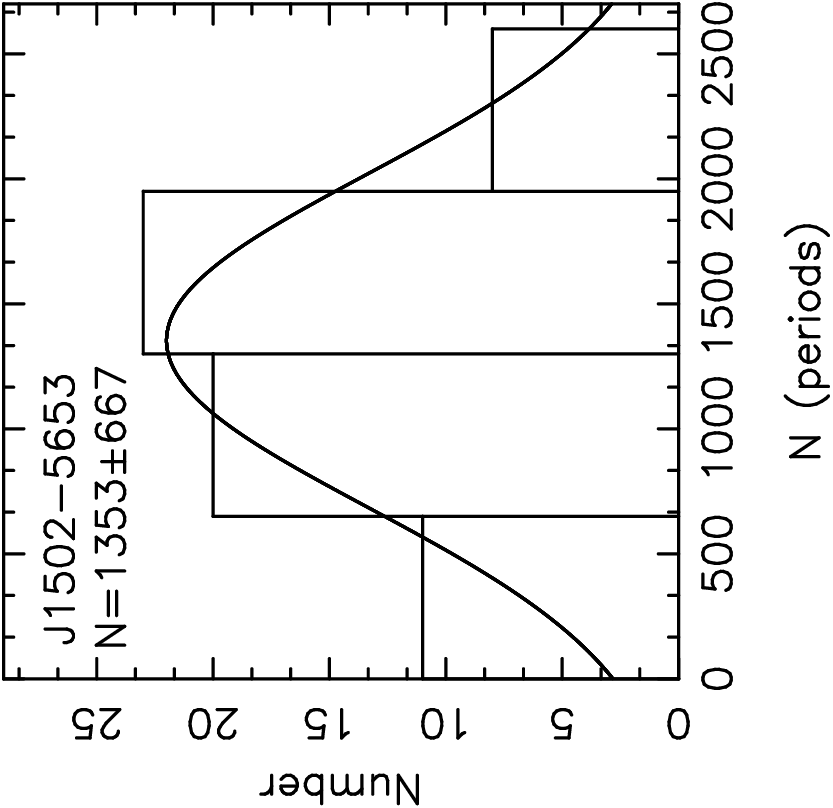}\\[2mm]
    \includegraphics[width = 0.30\textwidth, height=0.30\textwidth, angle = -90]{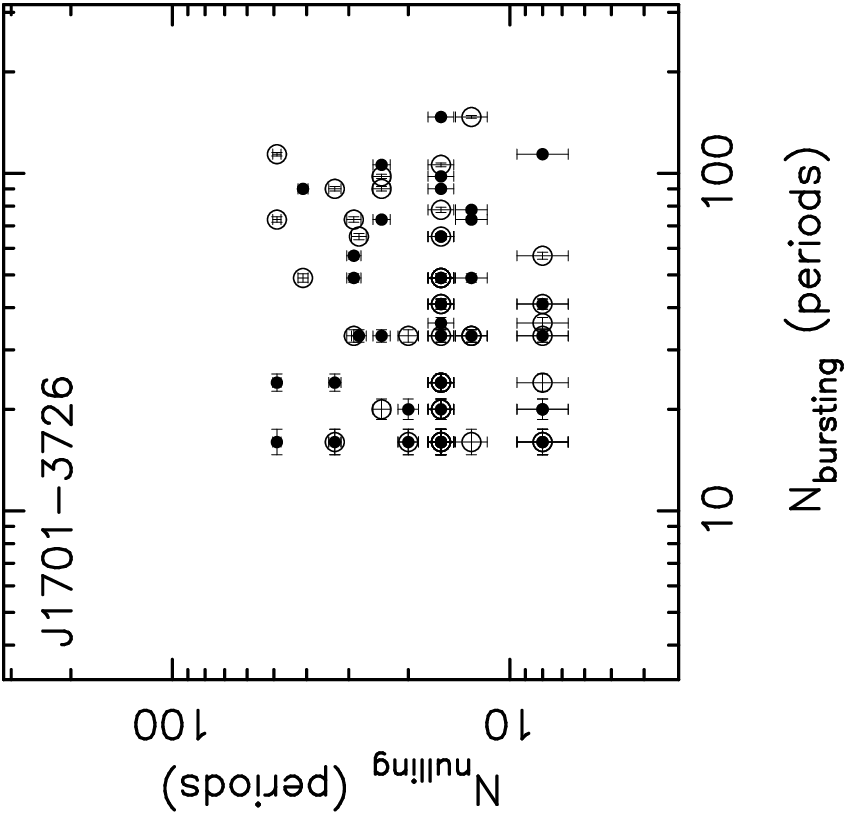}&
    \includegraphics[width = 0.30\textwidth, height=0.30\textwidth, angle = -90]{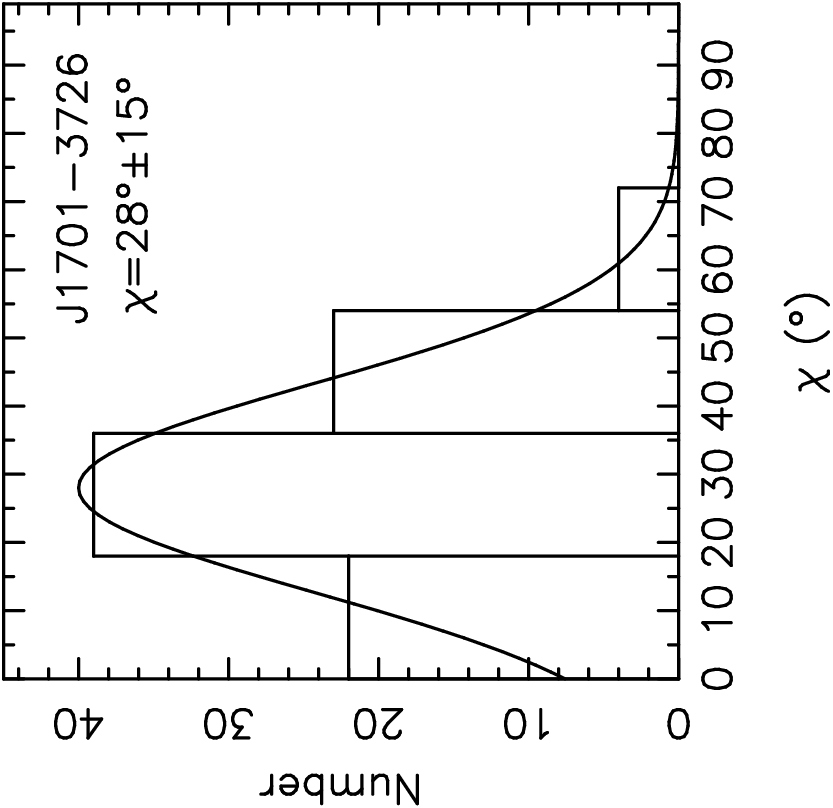}&
    \includegraphics[width = 0.30\textwidth, height=0.30\textwidth, angle = -90]{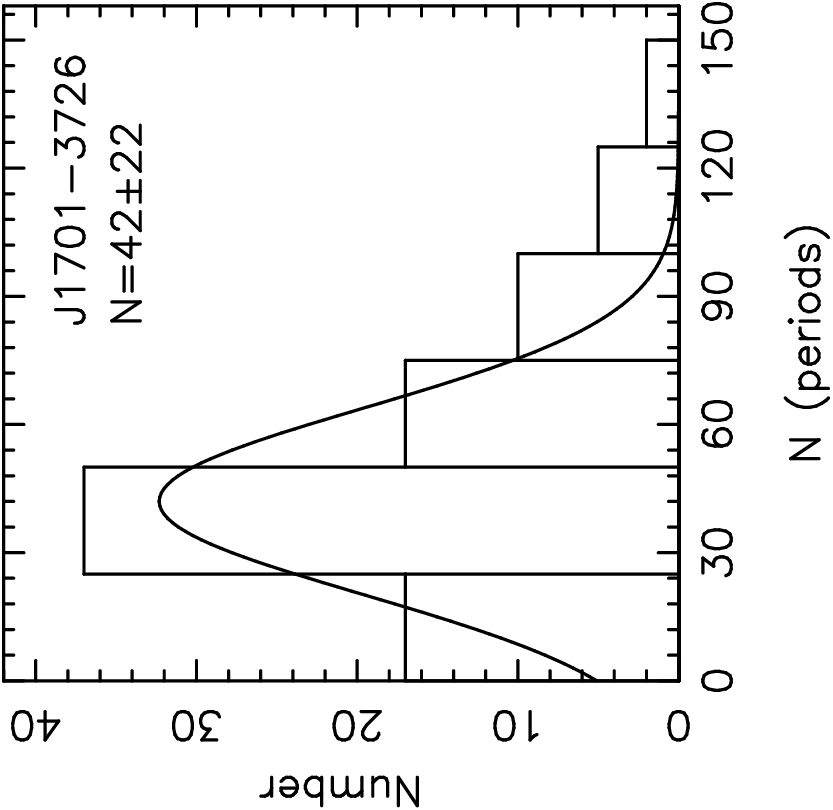}\\[2mm]
   \end{tabular}
  \caption{Distribution of period numbers of adjacent nulling and
    bursting ({\it left}), and the histogram distribution of nulling
    degree ({\it middle}) and nulling scale ({\it right}) and their
    Gaussian fittings. The black dots in the left plots stand for the
    nulling periods with the {\it subsequent} bursting period, and
    open circles for the nulling periods with their {\it preceding}
    bursting periods. Most data points in the left panels are measured
    from pulse sequences in literature. Some data points have similar
    $N_{\rm nulling}$ and/or $N_{\rm bursting}$, and hence are
    overlapped in plots in the logarithm-scale.}
  \label{fig2}
  \addtocounter{figure}{-1}
\end{figure*}
\begin{figure*}
  \begin{tabular}{ccc}
    \includegraphics[width = 0.30\textwidth, height=0.30\textwidth, angle = -90]{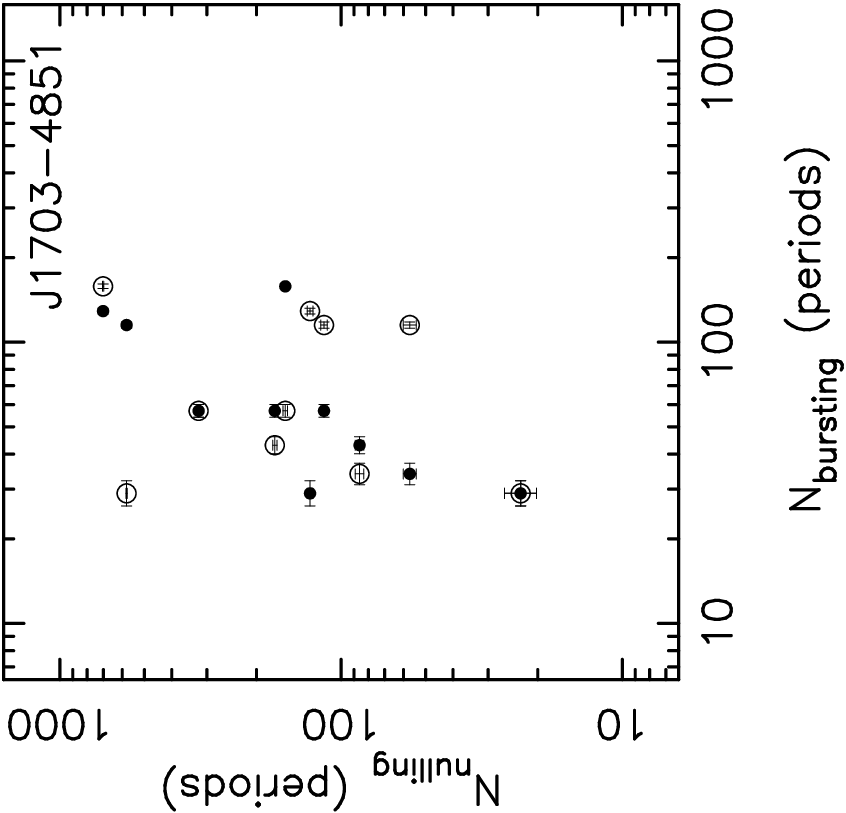}&
    \includegraphics[width = 0.30\textwidth, height=0.30\textwidth, angle = -90]{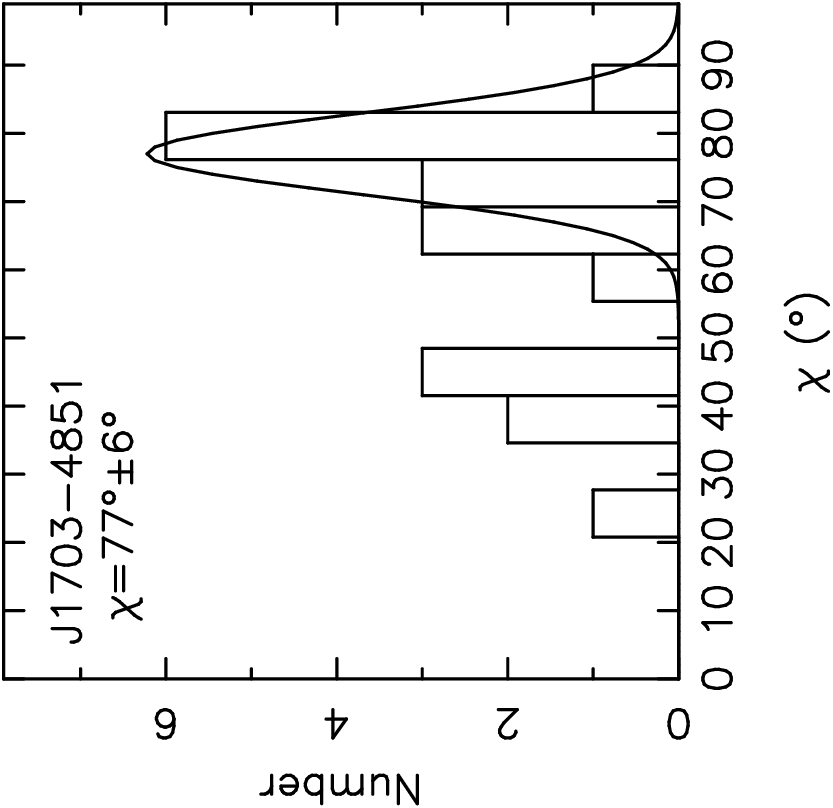}&
    \includegraphics[width = 0.30\textwidth, height=0.30\textwidth, angle = -90]{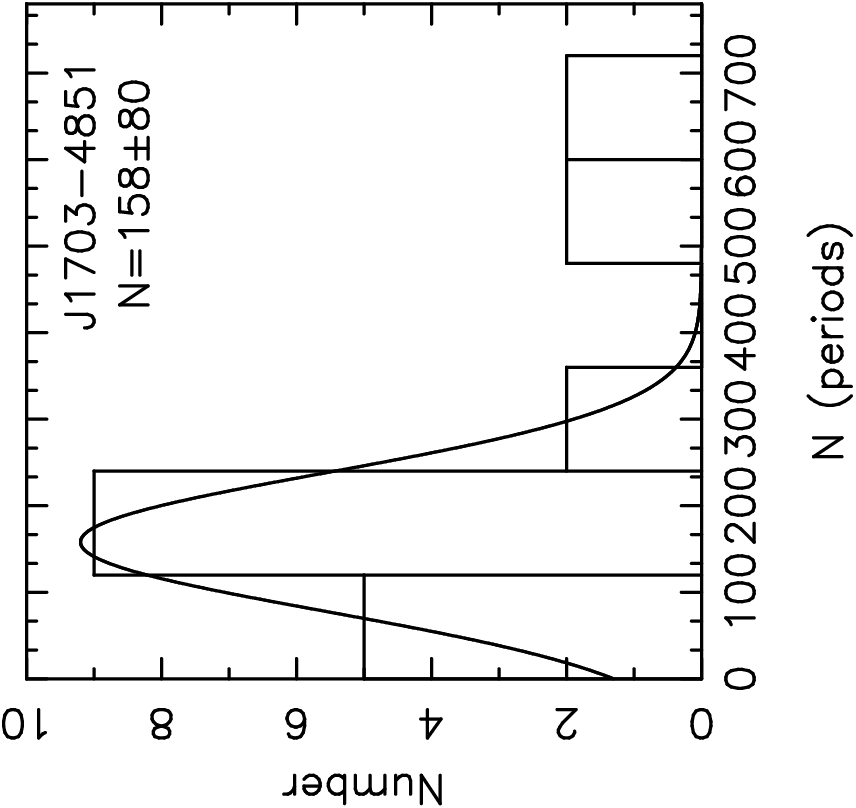}\\[2mm]
    \includegraphics[width = 0.30\textwidth, height=0.30\textwidth, angle = -90]{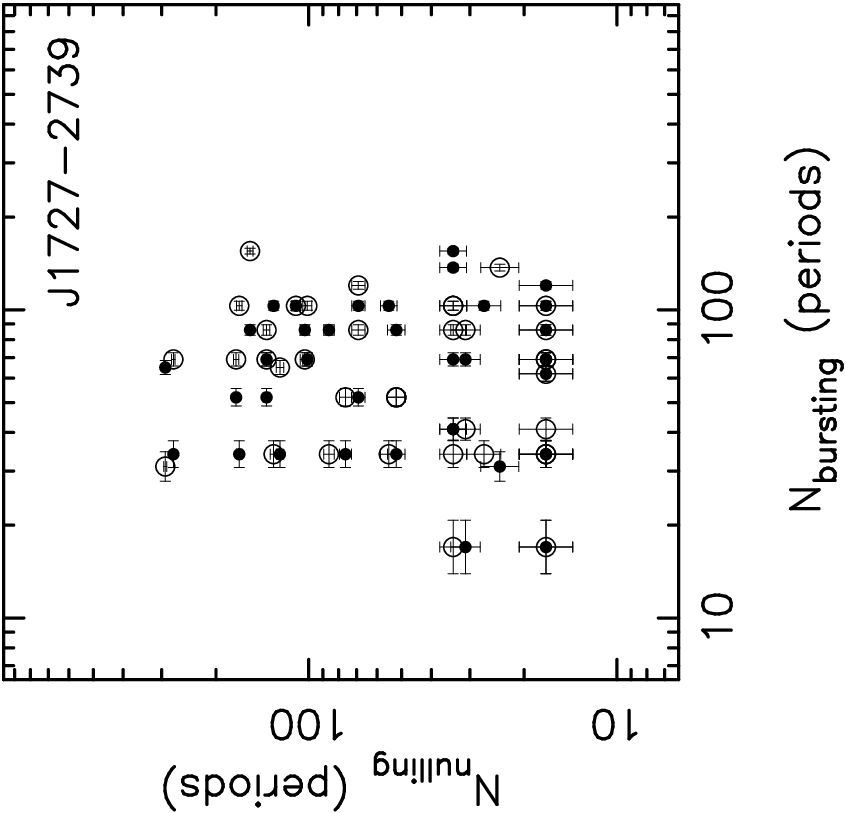}&
    \includegraphics[width = 0.30\textwidth, height=0.30\textwidth, angle = -90]{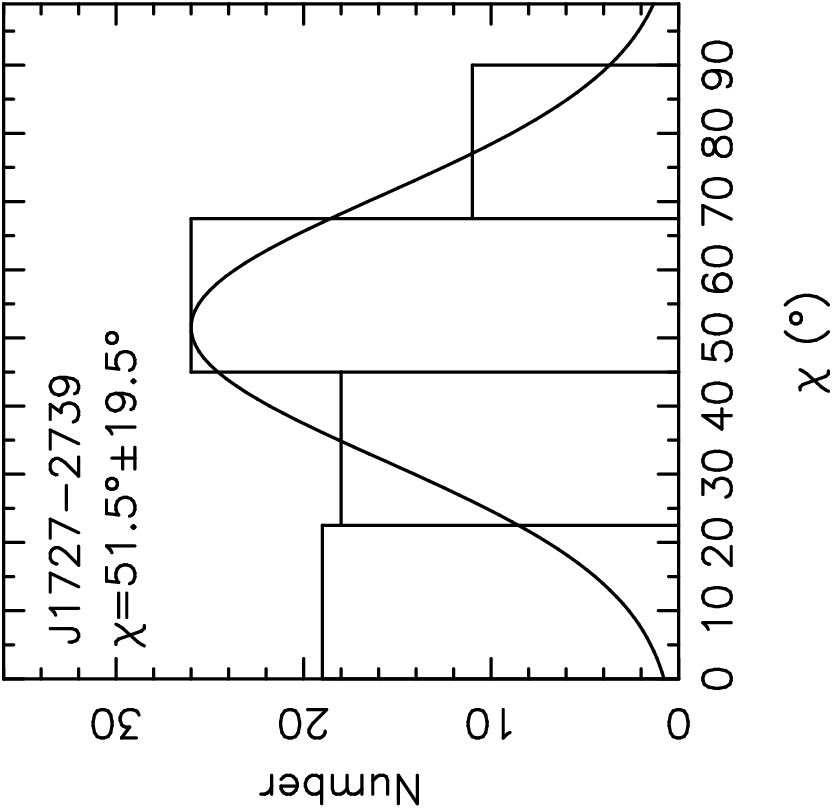}&
    \includegraphics[width = 0.30\textwidth, height=0.30\textwidth, angle = -90]{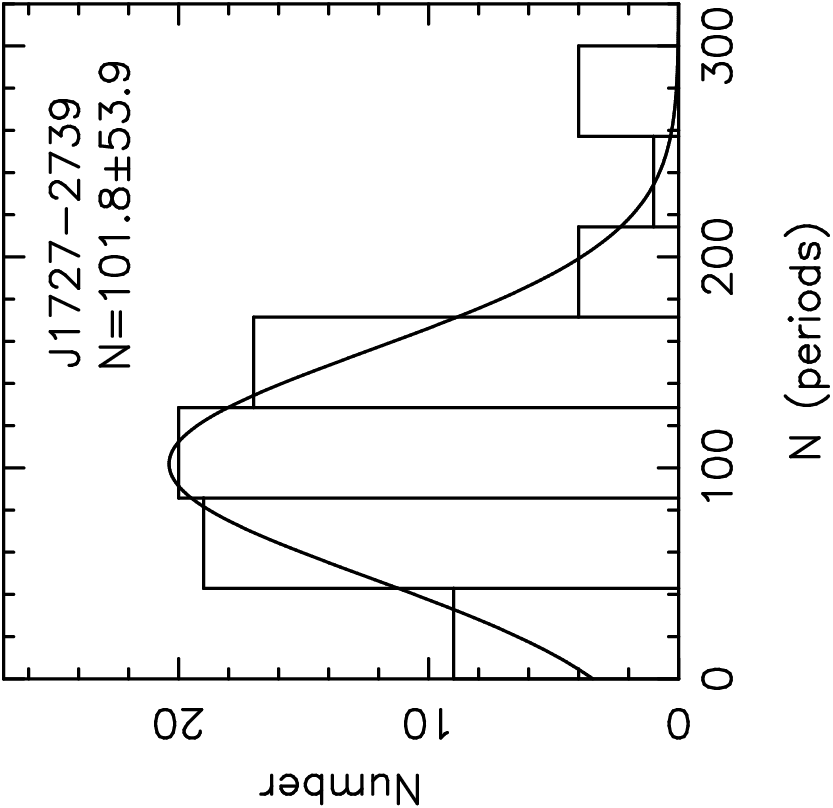}\\
    \includegraphics[width = 0.30\textwidth, height=0.30\textwidth, angle = -90]{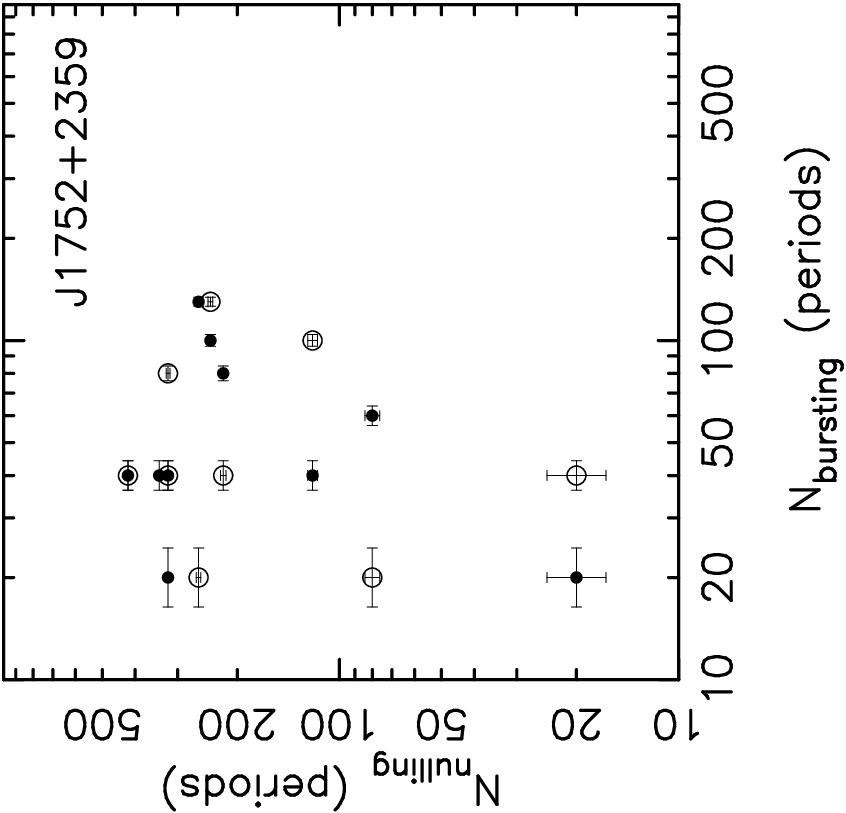}&
    \includegraphics[width = 0.30\textwidth, height=0.30\textwidth, angle = -90]{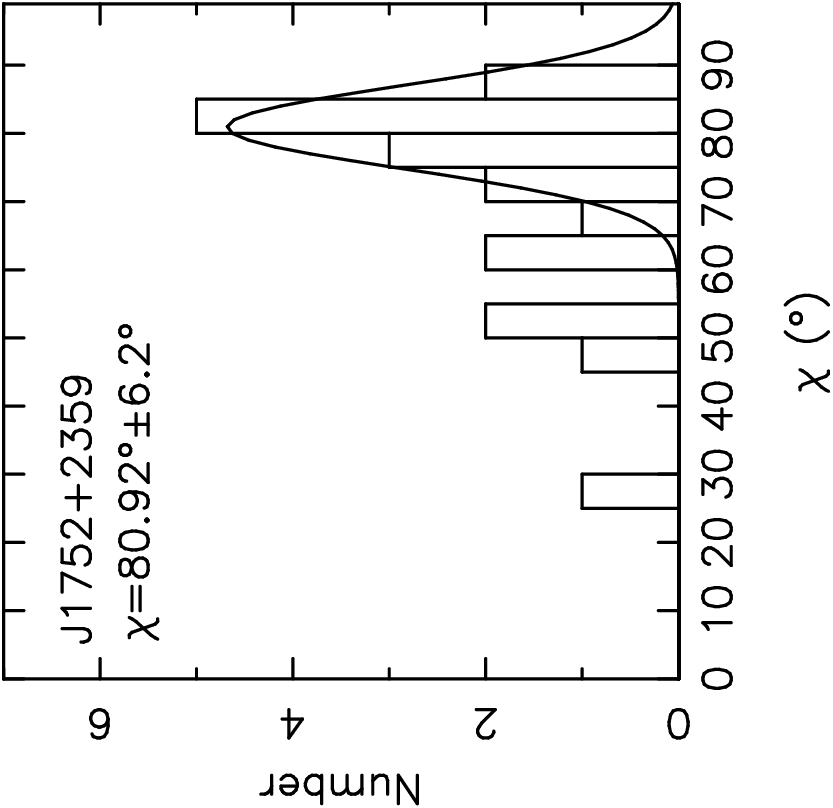}&
    \includegraphics[width = 0.30\textwidth, height=0.30\textwidth, angle = -90]{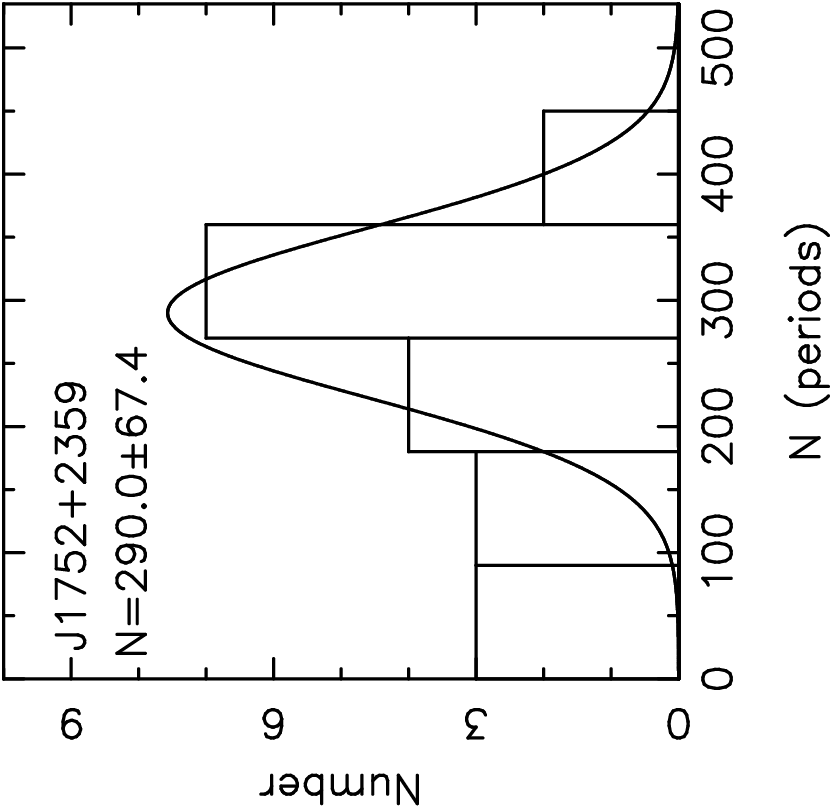}\\
    \includegraphics[width = 0.30\textwidth, height=0.30\textwidth, angle = -90]{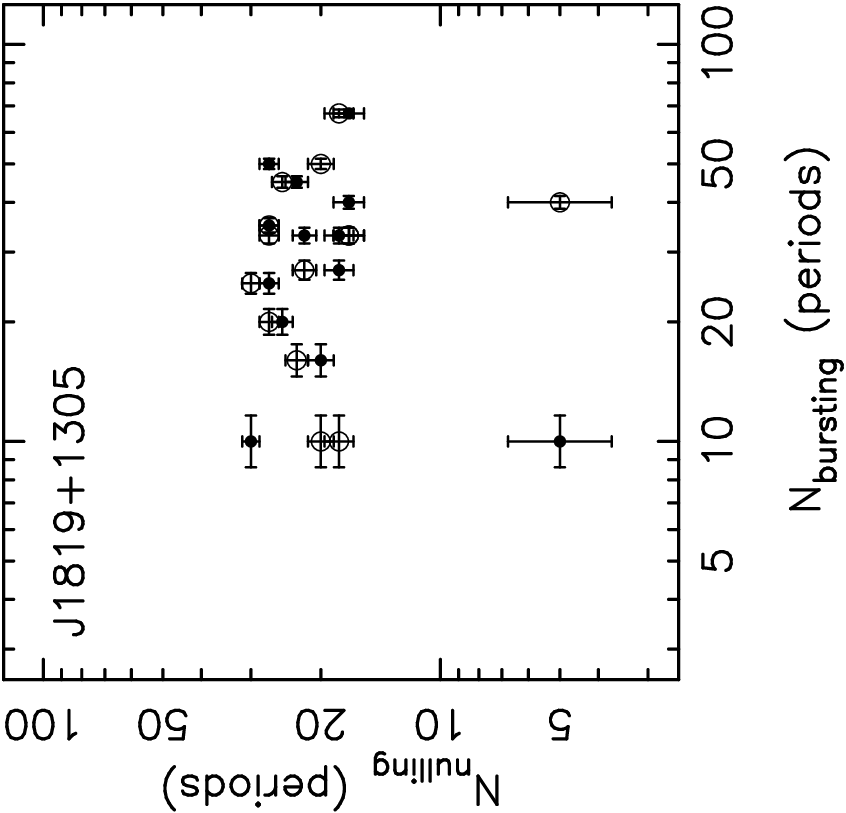}&
    \includegraphics[width = 0.30\textwidth, height=0.30\textwidth, angle = -90]{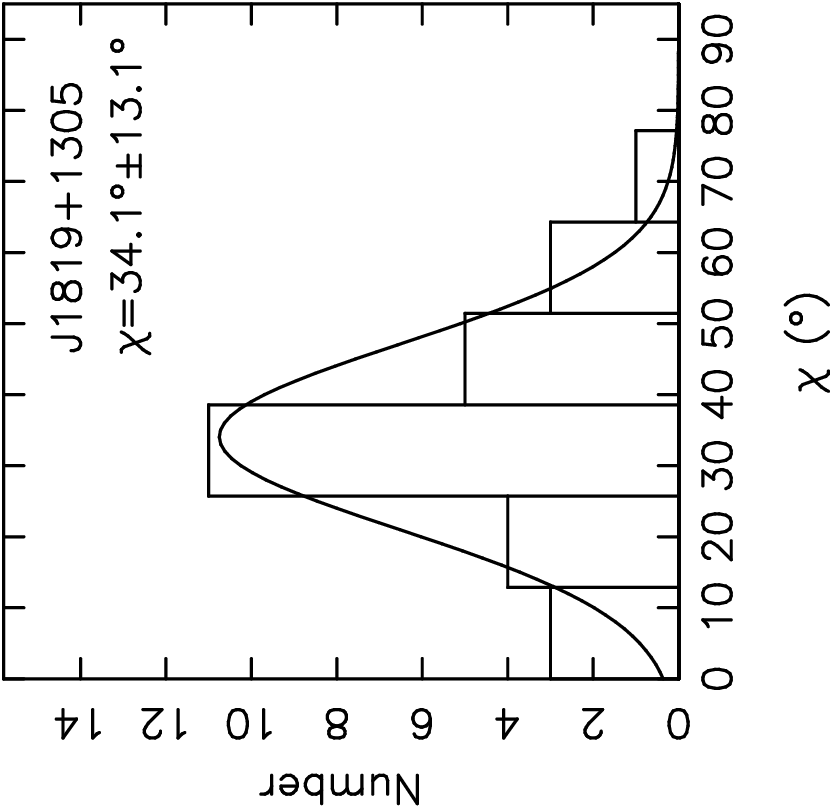}&
    \includegraphics[width = 0.30\textwidth, height=0.30\textwidth, angle = -90]{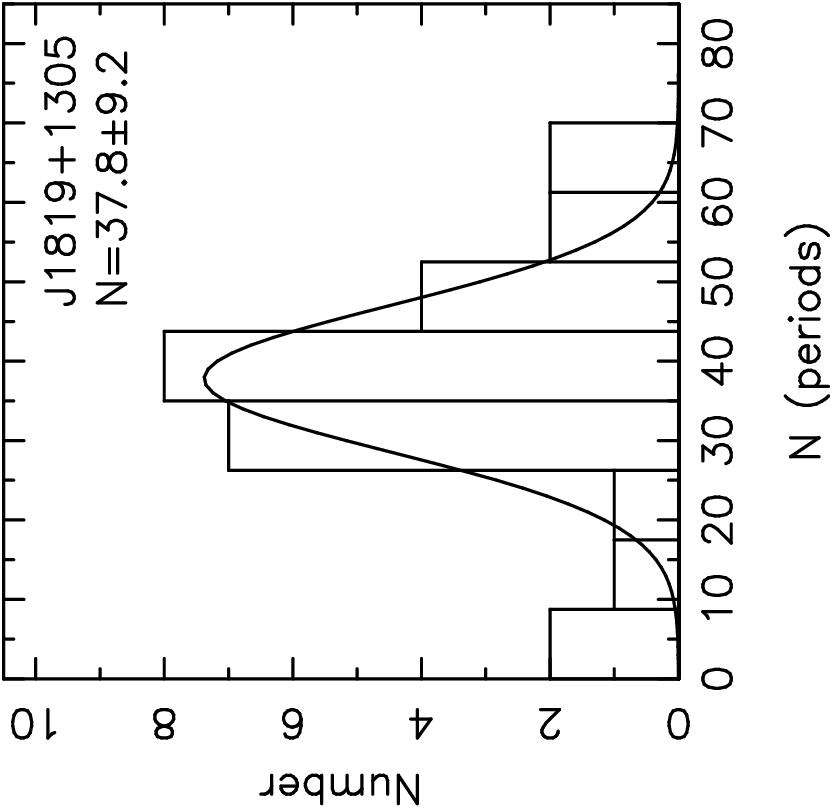}\\
   \end{tabular}
   \caption[]{-- {\it continued} }
   \addtocounter{figure}{-1}
   \end{figure*}
   \begin{figure*}
  \begin{tabular}{ccc}
    \includegraphics[width = 0.30\textwidth, height=0.30\textwidth, angle = -90]{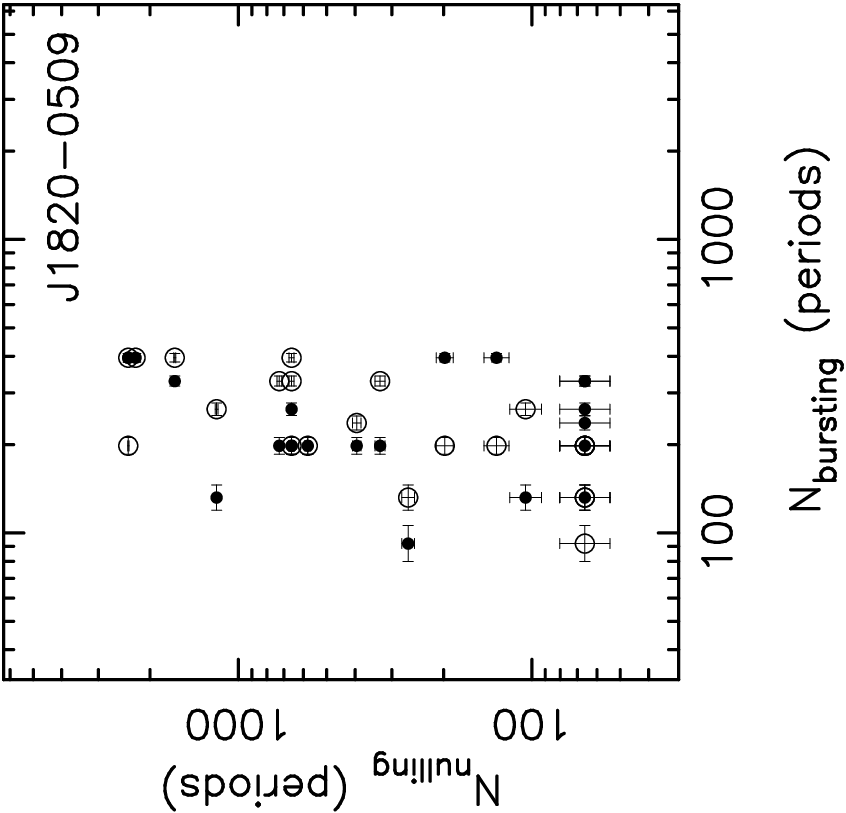}&
    \includegraphics[width = 0.30\textwidth, height=0.30\textwidth, angle = -90]{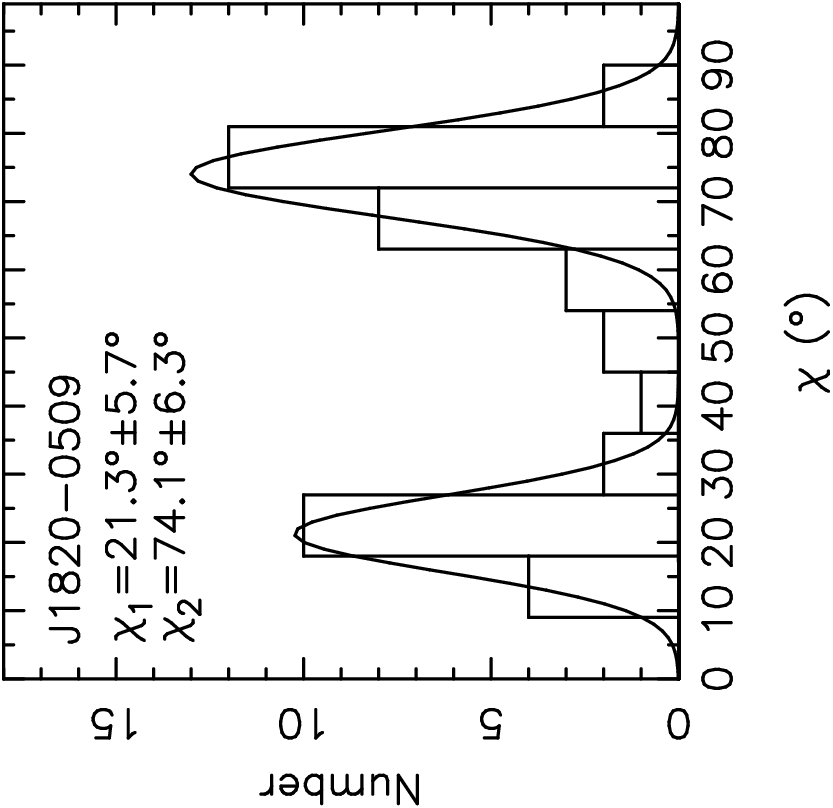}&
    \includegraphics[width = 0.30\textwidth, height=0.30\textwidth, angle = -90]{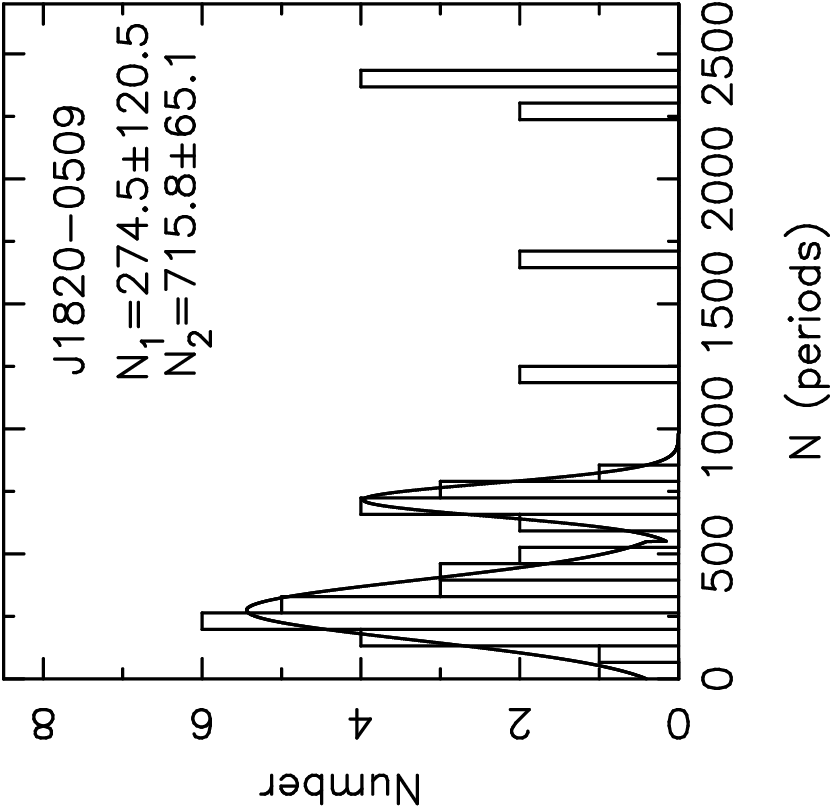}\\
    \includegraphics[width = 0.30\textwidth, height=0.30\textwidth, angle = -90]{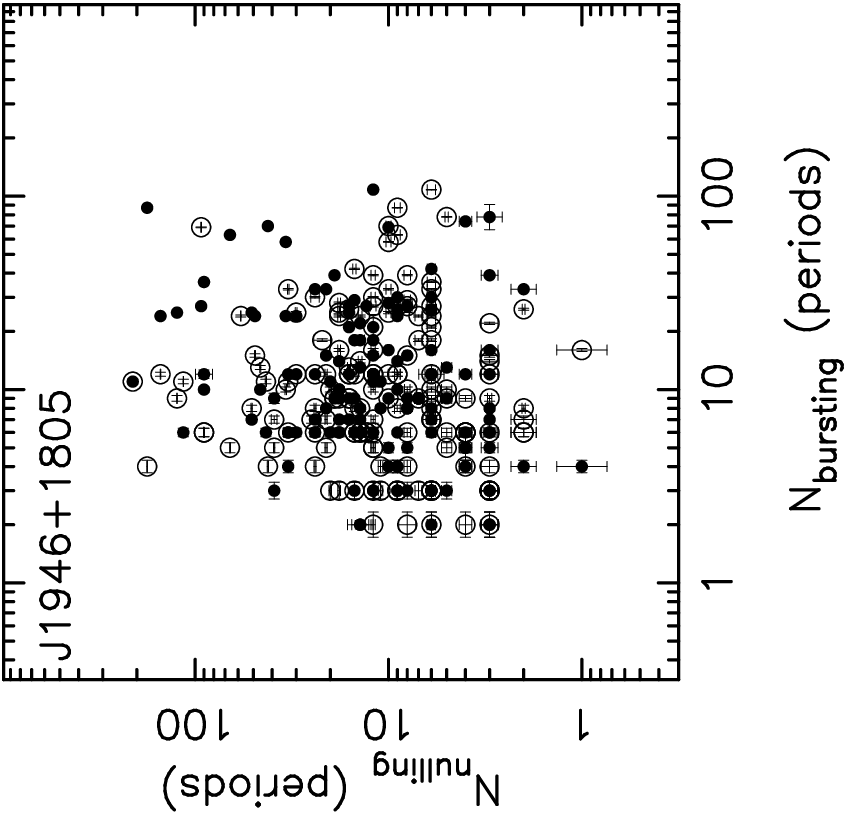}&
    \includegraphics[width = 0.30\textwidth, height=0.30\textwidth, angle = -90]{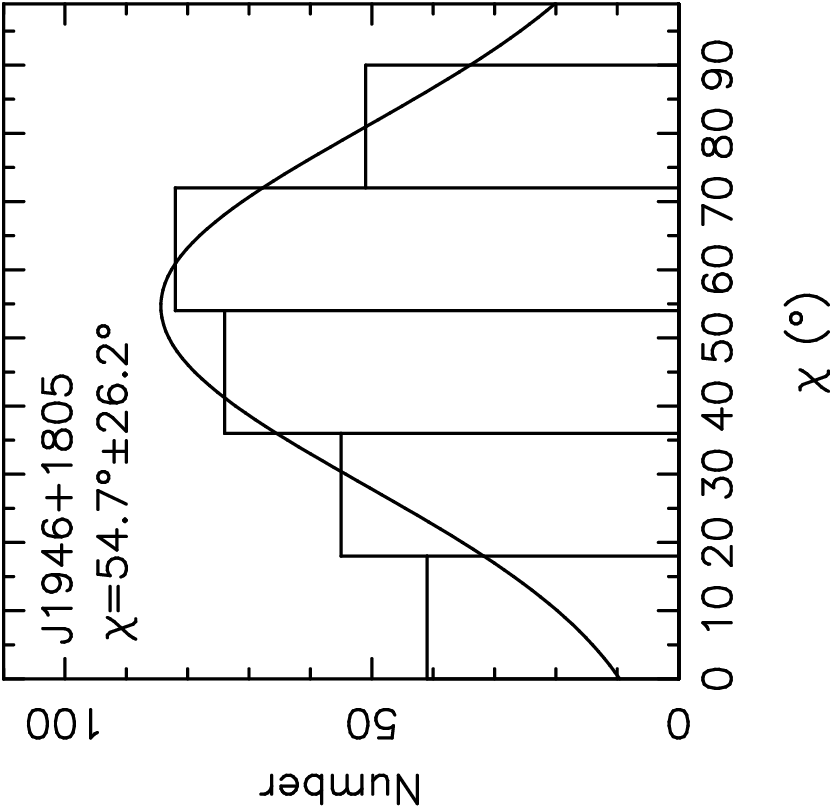}&
    \includegraphics[width = 0.30\textwidth, height=0.30\textwidth, angle = -90]{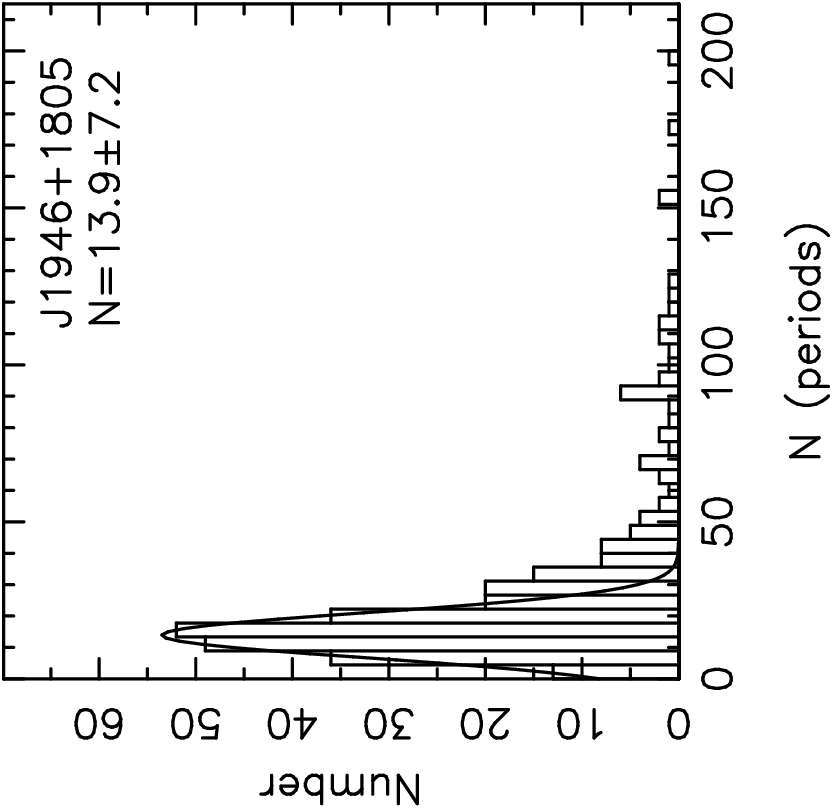}\\
 
 \end{tabular}
   \caption[]{-- {\it continued} } 
\end{figure*}

\begin{table*}
\small
\centering
\caption{Nulling degree and nulling scale for 10 pulsars, together
  with nulling fractions and nulling cycles ($Nc$) taken from
  literature. The pulsar parameters, the derivative of rotation
  frequency ($\nu'=\dot{P}/P^2$), the loss rate of rotational energy
  ($\dot{E}=4\pi I \dot{P}/P^3$) and the characteristic age
  ($\tau=P/2\dot{P}$) of pulsars are taken from the ATNF pulsar
  catalog[23].}
\begin{tabular}{ccccrrrr}
\\
\toprule[0.60pt]
 PSR-Jname  & $\chi$ & $ N$  & $NF$ & $Nc(s)$ & $\nu'$($10^{-16}$) & log $\dot{E}$ & log $\tau$ \\
   \hline
 J0820$-$4114& ${18}^{\circ}\pm 15^{\circ}$  & $47 \pm 20$ & 30     &                 &$-$0.64 &30.66 &8.66 \\  
 J1049$-$5833& ${54}^{\circ}\pm 14^{\circ}$  & $110\pm 24$ &$47\pm3$&$179\pm127$      &$-$9.1  &31.20 &6.90 \\ 
             & ${19}^{\circ}\pm\;7^{\circ}$  &  $46\pm 23$ &        &                 &        &      &     \\ 
 J1502$-$5653& ${81}^{\circ}\pm\;6^{\circ}$  &$1353\pm667$ &$93\pm4$&$515\pm360$      &$-$63.77&32.67 &6.67 \\ 
 J1701$-$3726& ${28}^{\circ}\pm\;15^{\circ}$  & $44 \pm\;22$ &$14\pm2$&$118\pm91 $      &$-$18.49&31.48 &6.54 \\
 J1703$-$4851& ${77}^{\circ}\pm\;6^{\circ}$  & $158 \pm80$ &74*    &                 &$-$26.05&31.87 &6.64 \\
 J1727$-$2739& ${52}^{\circ}\pm 20^{\circ}$  & $102\pm 54$ &$52\pm3$&$91\pm58$        &$-$6.58 &31.30 &7.27 \\ 
 J1752$+$2359& ${81}^{\circ}\pm\;6^{\circ}$  & $290\pm 67$ & 81     &                 &$-$38.41&32.57 &7.00 \\ 
 J1819$+$1305& ${34}^{\circ}\pm 13^{\circ}$  & $38 \pm\;9$ &$41\pm6$&                 &$-$3.19 &31.08 &7.67 \\ 
 J1820$-$0509& ${74}^{\circ}\pm\;6^{\circ}$  & $715\pm 65$ &$67\pm3$&$104\pm68$       &$-$81.93&32.98 &6.76 \\ 
             & ${21}^{\circ}\pm\;6^{\circ}$  & $275\pm121$ &        &                 &        &      &     \\ 
 J1946$+$1805& ${55}^{\circ}\pm 26^{\circ}$  & $14 \pm\;7$ &$55\pm5$&                 &$-$1.24 &31.04 &8.46 \\ 
\bottomrule[0.60pt]
\end{tabular} \\
* Corrected from literature by our measured data.\\
\label{table2}
\end{table*}

\begin{figure*}
\centering
\includegraphics[width = 0.33\textwidth,angle = -90]{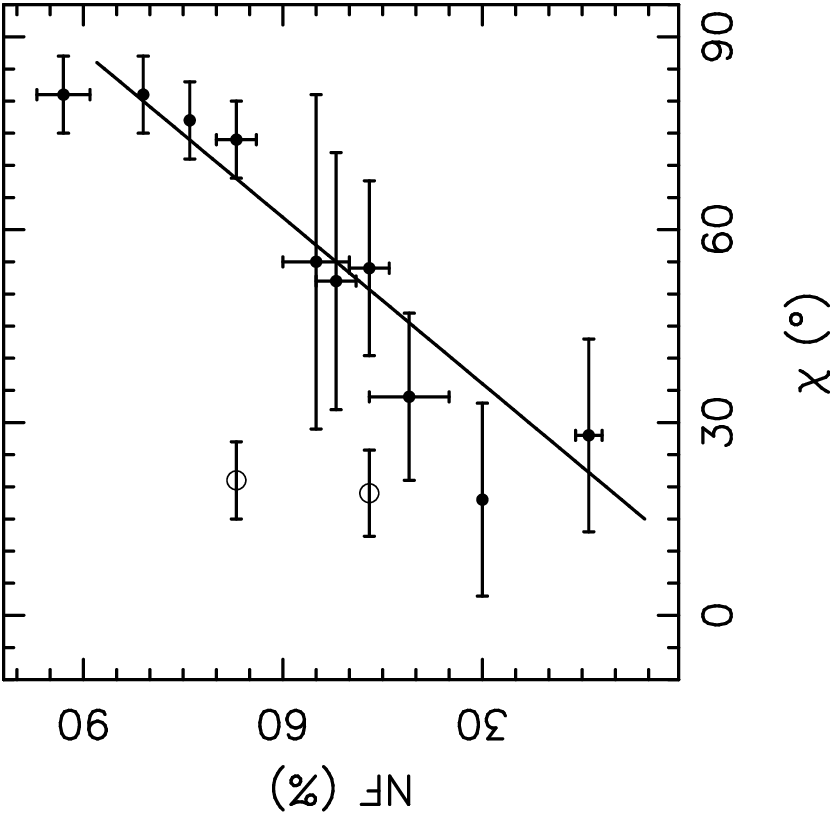}  \hspace{5mm}
\includegraphics[width = 0.33\textwidth,angle = -90]{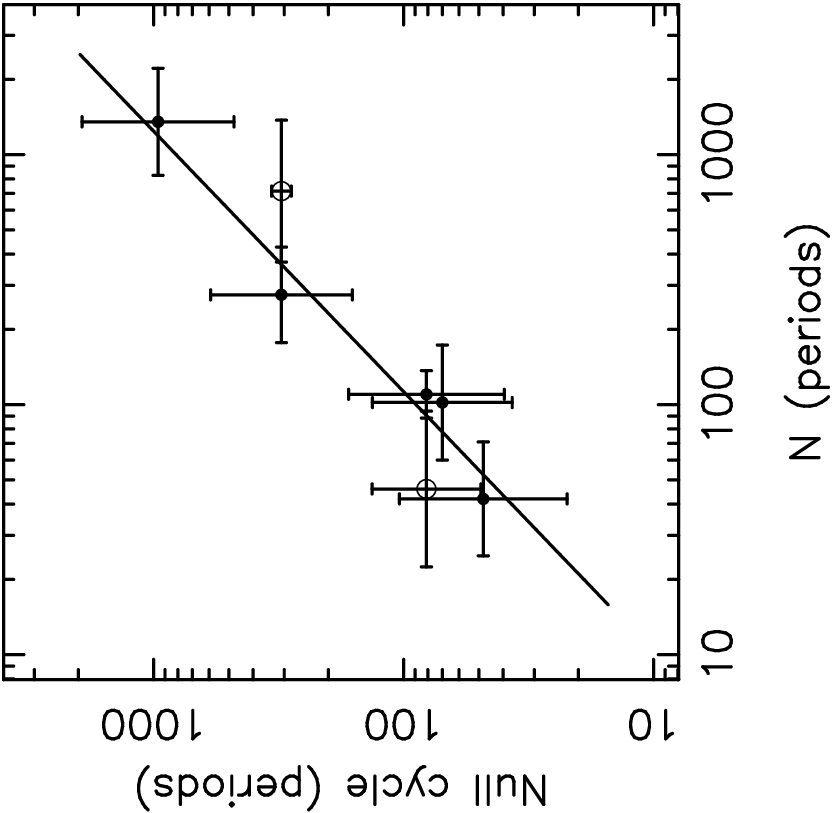}
\caption{{\it Left panel:} Strong correlation appears between the nulling degree
  ($\chi$) and the often-used nulling fraction (NF). The nulling
  degree value of PSR J1049$-$5833 and PSR J1820$-$0509 for the peak
  of the smaller amount of data is plotted as an open circle. {\it Right panel:}
  The null scale ($N$) is also found to be related to the nulling cycle in [5]. }
\label{fig3}
\end{figure*}

\begin{figure*}[t]
\centering
\includegraphics[width = 0.33\textwidth,angle = -90]{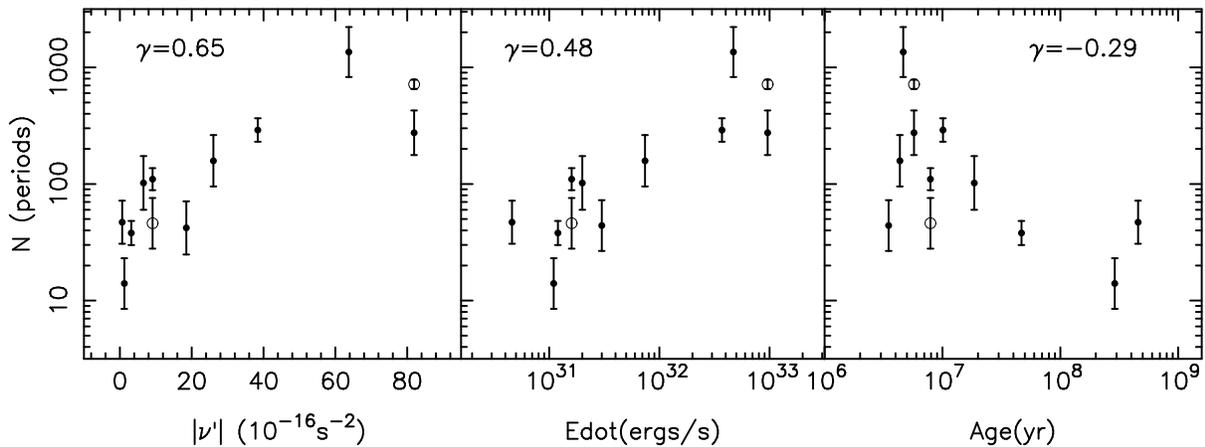}\vspace{-2mm}
\caption{Nulling scale ($N$) is obviously correlated to the derivative
  of rotation frequency ($\nu'=\dot{P}/P^2$) and the loss rate of
  rotational energy ($\dot{E}=4\pi I \dot{P}/P^3$), but not the
  characteristic age ($\tau=P/2\dot{P}$) of pulsars. The correlation
  parameter $\gamma$ is given in each panel. The values of PSR
  J1820$-$0509 and PSR J1049-5833 for the peak of the smaller amount
  of data are plotted with open circles.}
\label{fig4}
\end{figure*}

We collect observational nulling data from literature for 10 pulsars, 
as listed in Table 1, if there are sufficient data of pulse-sequences.
The nulling and bursting period numbers of PSR J0820$-$4114
(B0818$-$41) are taken from Table 1, 2, 3, 4 in [19], which were
observed by using the Giant Metrewave Radio Telescope (GMRT) at two
frequencies (325~MHz and 610~MHz) with bandwidth of 16 MHz. All data 
of these data are used in this work. 
The data of PSR J1049$-$5833, and also PSR J1701$-$3726, J1703$-$4851,
J1727$-$2739 and J1820$-$0509, are measured from Fig.2 and Fig.3 in
[5], which were observed by using Parkes 64-m telescope at 1518 MHz
with a bandwidth of 576 MHz. 
The data of PSR J1502$-$5653 are also measured from Fig.1 in [2]
which were observed by using the Parkes 64-m telescope at frequency of
1374 MHz with bandwidth of 288 MHz and also Fig.2 in [5] observed
at 1518 MHz with a bandwidth of 576~MHz. 
The pulse sequence data of PSR J1752$+$2359 are measured from Fig.5
observed with the Arecibo telescope at 430~MHz in [20]. 
The single-pulse data of PSR J1819$+$1305 are taken from Fig.4 in [11], 
observed at 327~MHz by the Arecibo Telescope. 
The single-pulse sequences of PSR J1946$+$1805 (B1944$+$17) are taken
from Fig.2 in [21] which were observed using Arecibo at 430~MHz. 

The distribution of period numbers of nulling and bursting for these
10 pulsars are plotted in Fig.2, and the histograms for $\chi$ and $N$
are fitted with a Gaussian function. Most of these $\chi$ and $N$
distributions show one peak, some narrow, some extended. The $\chi$
and $N$ distributions of PSR J1049$-$5833 and 1820$-$0509 have two
peaks. The mean and deviation of the $\chi$ and $N$ distributions of
10 pulsars are listed in Table 2, together with the nulling fractions
and nulling cycles ($Nc$) taken from literature and pulsar properties,
the derivative of rotation frequency, the loss rate of rotational
energy and the characteristic age of pulsars from the pulsar catalog
[23].

\sec{3\quad Discussion} 

The distributions of nulling degree $\chi$ and nulling scale $N$
exhibited a wide range of nulling behaviour, which provide insights
for the stability of emission region or the switches of emission modes
in pulsar magnetosphere.

As shown in Fig.3, the mean nulling degree $\chi$ is strongly
correlated with the often-used nulling fraction (NF) which is the
average percentage of nulling time. The larger the mean nulling
degree, the larger the nulling fraction. The distribution of nulling
degrees contains more information than the average percentage of
time. For example, random nulling length and bursting length should
give a very broad distribution of nulling degree, so that the
deviation of $\chi$ is large, more than 15 degree for PSR
J0820$-$4114, PSR J1727$-$2739 and PSR J1946$+$1805. About half
pulsars in our sample show a narrow distribution of a few degree
around one (for PSR J1049$-$5833 and J1820$-$0509 even two) prefered
nulling degree, which indicates the somehow periodical behaviour of
nulling.

The other key parameter is the nulling scale, $N$, which indicates how
often nulling occurs. The distributions of $N$ for these 10 pulsars
show a peak, often with a long tail, except PSR J1049$-$5833 and
J1820$-$0509 which show two main peaks. The nulling scale of PSR
J1502$-$5653, $N=1353\pm667$ periods (12$\pm$6 minutes) is consistent
with the quasi-periodicities of 11 minutes and 18 minutes found in
[2]. The mean of the nulling scale distribution is closely related to
the nulling cycle discussed in [5], as shown in Fig.4.  However, the
distributions of the nulling scale $N$ show more detailes of the
nulling process, which may be related to the emission process and
pulsar radiation pattern. We found in Fig.4 that the nulling scale $N$
is strongly related to the derivative of rotation frequency ${\nu}'$,
the loss rate of rotational energy ($\dot{E}=4\pi I \dot{P}/P^3$), but
not the characteristic age ($\tau=P/2\dot{P}$) of pulsars.

The standard deviation of the $N$ distributions also indicates the
randomness of nulling timescale. A large ${\sigma}_{N}$ for broad
distributions indicates random or non-periodic nullings. A narrow $N$
distributions can be used to predict the expected timescale of
bursting or nulling, which is a clue for
understanding pulsar radiation mechanism. For instance, when nulling
is caused by the empty passes of our sightline in the carousel-beam
system, the distribution of $\chi$ and $N$ can constrain the carousel
circulation time and the geometry of carousel-beam system.

 \Acknowledgements{\bahao We thank two anonymous referees for helpful
   comments and suggestions. The authors are supported by by China
   Ministry of Science and Technology under grant No. 2013CB837900,
   and the National Science Fundation of China under grant No.
   10833003, 11003023, 11273029, and 11261140641.}

\normalsize \vskip0.3in\parskip=0mm \baselineskip 18pt
\renewcommand{\baselinestretch}{1.1}\footnotesize\parindent=4mm\bahao

\REF{1\ } Backer D~C. Pulsar Nulling Phenomena. Nature, 1970, 228: 42--43

\REF{2\ } Li J, Esamdin A, Manchester R~N, Qian M~F, Niu H~B. 
Radiation properties of extreme nulling pulsar J1502-5653. Monthly Notices of the Royal Astronomical Society, 2012, 425: 1294--1298

\REF{3\ } Burke-Spolaor S, Bailes M. The millisecond radio sky: transients 
from a blind single-pulse search, Monthly Notices of the Royal Astronomical Society, 2010, 402: 855--866

\REF{4\ } Esamdin A, Lyne A~G, Graham-Smith F, et al. Mode switching and 
subpulse drifting in PSR B0826-34. Monthly Notices of the Royal Astronomical Society,
2005, 356: 59--65

\REF{5\ } Wang N, Manchester R~N, Johnston S. Pulsar nulling and mode 
changing. Monthly Notices of the Royal Astronomical Society, 2007, 377: 1383--1392

\REF{6\ } Lyne A~G, Ashworth M. The effect of nulls upon subpulse drift 
in PSRs 0809$+$74 and 0818$-$13. Monthly Notices of the Royal Astronomical Society, 1983, 204: 519--536

\REF{7\ } Unwin S~C, Readhead A~C~S, Wilkinson P~N, Ewing W~S. 
Phase stability in the drifting subpulse pattern of PSR 0809$+$74. Monthly Notices of the Royal Astronomical Society, 1978, 182: 711--721

\REF{8\ } Gil J~A, Jessner A, Kijak J, Kramer M, Malofeev V, Malov I, Seiradakis J H, Sieber W, Wielebinski R. Multifrequency study of PSR 1822-09. Astronomy and Astrophysics, 1994, 282: 45--53

\REF{9\ } Herfindal J~L, Rankin J~M. Periodic nulls in the pulsar 
B1133$+$16, Monthly Notices of the Royal Astronomical Society, 2007, 380: 430--436

\REF{10\ } Herfindal J~L, Rankin J~M. Deep analyses of nulling in Arecibo pulsars reveal
further periodic behaviour. Monthly Notices of the Royal Astronomical Society, 2009, 393: 1391--1402

\REF{11\ } Rankin J~M, Wright G~A~E. The periodic nulls of radio 
pulsar J1819$+$1305. Monthly Notices of the Royal Astronomical Society, 2008, 385, 1923--1930

\REF{12\ } Biggs J~D. An analysis of radio pulsar nulling statistics. 
Astrophysical Journal, 1992, 394: 574--580

\REF{13\ } Ritchings R~T. Pulsar single pulse intensity measurements 
and pulse nulling. Monthly Notices of the Royal Astronomical Society, 1976, 176: 249--263

\REF{14\ } Redman S~L, Rankin J~M. On the randomness of pulsar nulls. 
Monthly Notices of the Royal Astronomical Society, 2009, 395: 1529--1532

\REF{15\ } Vivekanand M. Observation of nulling in radio pulsars with 
the Ooty Radio Telescope. Monthly Notices of the Royal Astronomical Society, 1995, 274: 785--792

\REF{16\ } Kramer M, Lyne A~G, O'Brien J~T, Jordan C~A, Lorimer D~R. 
A Periodically Active Pulsar Giving Insight into Magnetospheric Physics, Science, 2006, 312: 549--551

\REF{17\ } Cordes J M. Pulsar State Switching from Markov Transitions and 
Stochastic Resonance. 2013, eprint arXiv:1304.5803

\REF{18\ } Gajjar V, Joshi B~C, Kramer M. A survey of nulling 
pulsars using the Giant Meterwave Radio Telescope, Monthly Notices of the Royal Astronomical Society, 2012, 424: 1197--1205

\REF{19\ } Bhattacharyya B, Gupta Y, Gil J.
Investigation of the unique nulling properties of PSR B0818$-$41, Monthly Notices of the Royal Astronomical Society, 2010, 408: 407C-421

\REF{20\ } Lewandowski W, Wolszczan A, Feiler G, Konacki M, Soltysinski T. 
Arecibo Timing and Single-Pulse Observations of Eighteen Pulsars. Astrophysical Journal, 2004, 600: 905--913

\REF{21\ } Deich W T S, Cordes J M, Hankins T H, Rankin J M. Null transition times, 
quantized drift modes, and no memory across nulls for PSR 1944 $+$ 17, Astrophysical Journal, 1986, 300: 540--550

\REF{22\ } Rankin J~M. Toward an empirical theory of pulsar emission. 
III - Mode changing, drifting subpulses, and pulse nulling. The Astronomical Journal, 1986, 301: 901--922

\REF{23\ } Manchester R N, Hobbs G B, Teoh A, Hobbs M. ATNF Pulsar Catalogue, 
Astronomical Journal, 2005, 129: 1993--2006 (http://www.atnf.csiro.au/research/pulsar/psrcat)

\label{lastpage}
\end{multicols}

\end{document}